\documentclass[12pt,a4paper,epsf]{article}
\usepackage{graphics}
\usepackage{amssymb,amsmath}
\usepackage[dvips]{lscape,graphicx}
\usepackage{cite}
\usepackage{longtable}
\newcommand{\ct}{\cite}
\newcommand{\lb}{\label}

\newcommand{\bc}{\begin{center}}
\newcommand{\ec}{\end{center}}
\newcommand{\bd}{\begin{displaymath}}
\newcommand{\ed}{\end{displaymath}}
\newcommand{\be}{\begin{equation}}
\newcommand{\ee}{\end{equation}}
\newcommand{\ba}{\begin{array}}
\newcommand{\ea}{\end{array}}
\newcommand{\bea}{\begin{eqnarray}}
\newcommand{\eea}{\end{eqnarray}}
\newcommand{\bt}{\begin{tabular}}
\newcommand{\et}{\end{tabular}}
\newcommand{\un}{\underline}

\newcommand{\bp}{\begin{picture}}
\newcommand{\ep}{\end{picture}}
\newcommand{\bfi}{\begin{figure}}
\newcommand{\efi}{\end{figure}}

\def\fun#1#2{\lower3.6pt\vbox{\baselineskip0pt\lineskip.9pt
\ialign{$\mathsurround=0pt#1\hfil##\hfil$\crcr#2\crcr\sim\crcr}}}
\begin{document}
\vspace{1cm}
\title{\LARGE \bf {False Vacuum Higgs Inflation and the Graviweak
Unification}}
\author{\large \bf  C.R.~Das${}^{1}$\footnote
{das@theor.jinr.ru}\,\, and L.V.~Laperashvili ${}^{2}$\footnote{laper@itep.ru}\\[5mm]
{\large \it ${}^{1}$ Institute of Physics,}\\
{\large \it Sachivalaya Marg, Bhubaneswar - 751005, Odisha, India}\\\\
{\large \it ${}^{2}$ The Institute of Theoretical and
Experimental Physics,}\\
{\large\it National Research Center ``Kurchatov Institute",}\\ {\large
\it Bolshaya Cheremushkinskaya, 25, 117218 Moscow, Russia}}
\date{}
\maketitle
\thispagestyle{empty}

\begin{abstract}

In the present paper we develop a model of the Higgs inflation
based on the non-minimal coupling of the Higgs boson to gravity
predicted by Graviweak Unification. But later we get rid of the
non-minimal coupling to gravity by making the conformal
transformation from the Jordan frame to the Einstein frame. We
construct a self-consistent $Spin(4,4)$-invariant model of the
unification of gravity with weak $SU(2)$ interactions in the
assumption of the existence of visible and invisible sectors of
the Universe. Assuming the interaction between the ordinary and
mirror Higgs fields, we develop a special Hybrid model of
inflation. According to this model, the inflaton starts trapped
from the "false vacuum" of the Universe at the Higgs field VEV
$v\sim 10^{18}$ GeV (in the visible world). Then the inflations of
the two Higgs doublet fields, visible $\phi$ and mirror $\phi'$,
lead to the emergence of the Standard Model vacua at the
Electroweak scales with the Higgs boson VEVs $v_1\approx 246$ GeV
and $v'_1=\zeta v_1$  in the visible and invisible worlds,
respectively. Considering the results of cosmology and calculating
the number of e-folds $N^*$, we predict $\zeta\simeq 100-115$ for
$N^*\simeq 50-60$, in agreement with previous results of the model
with  broken mirror parity. We also consider RGEs, taking into
account the mixing term - the interaction between the ordinary and
mirror Higgs bosons, assuming the smallness of the mixing
coupling.

\end{abstract}

\vspace{1cm}


\par cosmological constant, dark energy


\thispagestyle{empty}

\clearpage\newpage

\section{Introduction}

In Ref.~\ct{LNS} (see also \ct{1a,2a}) we suggested  a model of
the Higgs inflation, which follows from the unification of the
gravity, weak $SU(2)$ gauge and Higgs fields \ct{1,2}. In this
model we also used the Sidharth's ideas about the existence of a
discrete space-time at the Planck scale with non-commutative
geometry, predicting an almost zero cosmological constant
\ct{1S,2S,3S}.

Recently there has been a lot of interest in using the Higgs boson
as an inflaton in the context of the non-minimal coupling to
gravity \ct{1Bez,2Bez,3Bez,4Bez,5Bez,6Bez,1B,2B,3B,4B,5B,FGH}. The
papers \ct{FGH,1FV,2FV,3FV,4FV,5FV} investigated how the false
vacuum could be used to explain the inflationary phase of the
Universe. Inflation from a local minimum develops a model with a
graceful exit to the radiation-dominated era of the Universe. The
hypothesis that the inflation took place in the SM false vacuum is
consistent with a narrow range of values of the Higgs boson mass,
which subsequently turned out to be compatible with the
experimental range indicated by ATLAS and CMS \ct{1Hig,2Hig}.

In the present paper we developed the model of the Higgs inflation
using the Graviweak Unification \ct{1}, which predicts the
non-minimal coupling to gravity. We tried to calculate the
parameters of theory, which lead to the self-consistent
inflationary model. Assuming the interaction between the ordinary
and mirror Higgs fields $\varphi$ and $\varphi'$, we considered a
special Hybrid model of the inflation by A. Linde \ct{Lin}.
According to this inflationary model, a scalar field (inflaton)
starts trapped from the "false vacuum" of the Universe at the
Higgs field's VEV $v\sim 10^{18}$ GeV.

A model of unification of gravity with the weak $SU(2)$ gauge and
Higgs fields was constructed in Ref.~\ct{1}, in accordance with
Ref.\ct{2}. Previously gravi-weak and gravi-electro-weak unified
models were suggested in Refs.~\ct{1gw,2gw,3gw}.

In the present investigation we imagine that at the early stage of
the evolution of the Universe the GUT-group was broken down to the
direct product of gauge groups of the internal symmetry $U(4)$ and
$Spin(4,4)$-group of the Graviweak Unification.

In the assumption that there exist visible and invisible (hidden)
sectors of the Universe, we presented the hidden world as a Mirror
World (MW) with a broken Mirror Parity (MP), and gave arguments
that MW is not identical to the visible Ordinary World (OW). We
started with an extended $\mathfrak g =
\mathfrak{spin}(4,4)_L$-invariant Plebanski action in the visible
Universe, and with $\mathfrak g =
\mathfrak{spin}(4,4)_R$-invariant Plebanski action in the MW. Then
we have shown that the Graviweak symmetry breaking leads to the
following sub-algebras: $\tilde {\mathfrak g} = {\mathfrak
sl}(2,C)^{(grav)}_L \oplus {\mathfrak su}(2)_L$ -- in the ordinary
world, and $\tilde {\mathfrak g}' = {{\mathfrak
sl}(2,C)'}^{(grav)}_R \oplus {\mathfrak su}(2)'_R$ -- in the
hidden world. These sub-algebras contain the self-dual left-handed
gravity in the OW, and the anti-self-dual right-handed gravity in
the MW. Finally, at low energies, we obtain a Standard Model (SM)
group of symmetry and the Einstein-Hilbert's gravity. In this
approach we have developed a model of inflation, in which the
inflaton $\sigma$, being a scalar $SU(2)$-triplet field, decays
into the two Higgs $SU(2)$ doublets of the SM: $\sigma\to
\phi^\dagger \phi$, and then the interaction between the ordinary
and mirror Higgs fields (induced by gravity) leads to the hybrid
model of the inflation.

In Section 2 we considered the Plebanski's theory of gravity, in
which fundamental fields are 2-forms, containing tetrads, spin
connections and auxiliary fields. Then we have used an extension
of the Plebanski's formalism of the 4-dimensional gravitational
theory, and in Section 3 we constructed the action of the
Graviweak unification model, described by the overall unification
parameter $g_{uni}$. The existence of de Sitter solutions at the
early time of acceleration era of the Universe is discussed in
Subsection 3.1. The parameters of the Graviweak unification model
(Newton gravitational constant, $G_N$, bare cosmological constant,
$\Lambda_0$, and bare coupling constant of weak interaction,
$g_W$) were estimated in Subsection 3.2. It was shown that the GWU
Lagrangian includes the non-minimal coupling with gravity. We see
that at the Planck scale "second minimum", the Higgs field
$\varphi$ can be represented as $\varphi = v - \sigma$, where the
real scalar field $\sigma$ is an inflaton. At the Planck scale
vacuum with its VEV equal to $v$, the inflaton field is zero
($\sigma = 0$), and then increases with the falling of the field
$\varphi$. Considering the expansion of GWU Lagrangian in powers
of small values of $\sigma/v$, we get rid of the non-minimal
coupling to gravity by making the conformal transformation from
the Jordan frame to the Einstein frame. Section 4 is devoted to
the Multiple Point Model (MPM), which allows the existence of
several minima of the Higgs effective potential with the same
energy density (degenerate vacua). The MPM assumes the existence
of the SM itself up to the scale $\sim 10^{18}$ GeV, and predicts
that there exist two degenerate vacua into the SM: the first one
-- at the Electroweak (EW) scale (with the VEV $v_1\simeq 246$
GeV), and the second one -- at the Planck scale (with the VEV
$v=v_2\sim 10^{18}$ GeV). In Section 5 we considered the existence
in the Nature of the Mirror World (MW) with a broken Mirror Parity
(MP): the Higgs VEVs of the visible and invisible worlds are not
equal,{\footnote {In this paper the superscript 'prime' denotes
the M- or hidden H-world.} $\langle\phi\rangle=v$,
$\langle\phi'\rangle=v'$ and $v\neq v'$. The parameter
characterizing the violation of the MP is $\zeta = {v'}/{v} \gg
1$. In Section 6 we have used the Sidharth's prediction about the
existence of a discrete space-time at the Planck scale, showing
that the idea of non-commutativity predicts an almost zero
cosmological constant. In Section 7 we suggest a model of the
Higgs inflation using the GWU action with Lagrangian in the
Einstein frame. Taking into account the interaction between the
initial ordinary and mirror Higgs fields: $\alpha_\varphi
(\varphi^{\dagger} \varphi)( {\varphi'} ^{\dagger}{\varphi'})$, we
constructed a hybrid model of the Higgs inflation in the Universe.
According to this model, a scalar field $\varphi$ starts trapped
from the "false vacuum" of the Universe at the value of the Higgs
field's VEV $v =v_2 \sim 10^{18}$ GeV. Using new fields
$\varphi=v-\sigma$ and $\varphi'=v'-\sigma'$, we assume that a
scalar field $\sigma$, being an inflaton, starts trapped from the
"false vacuum" of the Universe at the value of the Higgs field's
VEV $v =v_2 \sim 10^{18}$ GeV. But then during inflation the field
$\sigma$ decays into the two Higgs doublets of the SM: $\sigma\to
\phi^\dagger \phi$. Considering the interaction $\alpha_\phi
(\phi^{\dagger} \phi)( {\phi'} ^{\dagger}{\phi'})$ between the
visible and mirror Higgs doublet fields $\phi$ and $\phi'$ , we
show that this interaction leads to the emergence of the SM vacua
at the EW scales with the Higgs boson VEVs $v_1\approx 246$ GeV
and $v'_1=\zeta v_1$ in the visible and invisible worlds,
respectively. Here we also use the Sidharth's prediction about the
existence of the non-commutative geometry at the Planck scale
which predicts an almost zero cosmological constant. In Subsection
7.1 we investigate the agreement of our GWU model of the Higgs
inflation with modern predictions of cosmology. Calculating the
expression for a number of e-folds $N^*$, we estimate the MW
parameter $\zeta$. We obtained: $\zeta \simeq 100-115\quad
{\rm{for}} \quad N^* \simeq 50-60,$ in accordance with estimations
of the previous references, predicted $\zeta\sim 100$. In Section
8 we presented the calculation of the renormalization group
equations (RGEs) in the assumption that there exists the
interaction between the ordinary and mirror Higgs bosons. We
confirmed the small values of parameters $\lambda'$ and
$\alpha_\phi$, which do not change essentially the results of the
2-, or 3-loop calculations of the Higgs mass. Section 9 contains
Summary and Conclusions.

\section{Plebanski's formulation of General Relativity}

General Theory of Relativity (GTR) was formulated by Einstein as
dynamics of the metrics  $g_{\mu\nu }$. Later, Plebanski \ct{3}
and other authors (see for example \ct{3a,3b}) presented GTR in
the self-dual approach, in which fundamental variables are 1-forms
of connections $A^ {IJ}$ and tetrads $e^I$ :
\be   A^{IJ} =  A_{\mu}^{IJ}dx^{\mu}, \qquad
       e^I = e_{\mu}^Idx^{\mu}.
                          \lb{1} \ee
Also 1-form $A = \frac 12 A^ {IJ}\gamma_ {IJ}$ is used, in which
generators $\gamma_ {IJ}$ are products of generators of the
Clifford algebra $Cl(1,3)$: $\gamma_ {IJ} = \gamma_{I}\gamma_
{J}$. Indices $I, J=0,1,2,3$ belong to the spacetime with
Minkowski's metrics $\eta^ {IJ} = {\rm diag} (1,-1,-1,-1)$, which
is considered as a flat space, tangential to the curved space with
the metrics $g_{\mu\nu}$. In this case connection belongs to the
local Lorentz group  $SO(1,3)$, or to the spin group $Spin(1,3)$.
In general case of unifications of gravity with the $SU (N)$ or
$SO (N)$ gauge and Higgs fields (see \ct{2}), the gauge algebra is
$\mathfrak g = \mathfrak {spin}(p, q)$, and we have $I, J = 1,2...
p+q$. In our model of unification of gravity with the weak $SU(2)$
interactions we consider a group of symmetry with the Lie algebra
$\mathfrak {spin}(4,4)$. In this model indices $I,J$ run over all
$8\times 8$ values: $I, J = 1,2..,7,8$.

For the purpose of construction of the action for any unification
theory, the following 2-forms are also considered:
$$ B^{IJ} = e^I\wedge e^J = \frac 12
e_{\mu}^Ie_{\nu}^Jdx^{\mu}\wedge dx^{\nu},\qquad F^{IJ} = \frac 12
F_{\mu\nu}^{IJ}dx^{\mu}\wedge dx^{\nu}, $$ where $F_{\mu\nu}^{IJ}
=\partial_{\mu}A_{\nu}^{IJ} - \partial_{\nu}A_{\mu}^{IJ} +
\left[A_{\mu}, A_{\nu}\right]^{IJ}$, which determines the
Riemann-Cartan curvature: $R_{\kappa \lambda \mu \nu} =
e_{\kappa}^I e_{\lambda}^JF_{\mu\nu}^{IJ}$. Also 2-forms of $B$
and $F$ are considered :
\be B= \frac 12 B^{IJ}\gamma_{IJ}, \qquad  F= \frac 12
F^{IJ}\gamma_{IJ}, \qquad  F = dA + \frac 12 \left[A, A\right].
\lb{9} \ee
The well-known in literature Plebanski's $BF$-theory is submitted
by the following gravitational action with nonzero cosmological
constant $\Lambda$:
\be  I_{(GR)} = \frac{1}{\kappa^2}\int
\epsilon^{IJKL}\left(B^{IJ}\wedge
    F^{KL} + \frac{\Lambda}{4}B^{IJ}\wedge B^{KL}\right),
                                  \lb{11} \ee
where $\kappa^2=8\pi G_N$, $G_N$ is the Newton's gravitational
constant, and $M_{Pl}^{red.} = 1/{\sqrt{8\pi G_N}}$.

Considering the dual tensors: $$F^*_{\mu\nu}\equiv \frac
{1}{2\sqrt{-g}}\epsilon^{\rho\sigma}_{\mu\nu}F_{\rho\sigma}, \quad
A^{\star IJ} = \frac 12 \epsilon^{IJKL}A^{KL},$$ we can determine
self-dual (+) and anti-self-dual (-) components of the tensor $A^
{IJ}$:
\be A^{(\pm)\,IJ}=\left({\cal P}^{\pm}A\right)^{IJ} = \frac 12
\left(A^{IJ} \pm iA^{\star\,IJ}\right).
                                 \lb{14} \ee
Two projectors on the spaces of the so-called self- and
anti-self-dual tensors $${\cal P}^{\pm}= \frac
12\left(\delta^{IJ}_{KL} \pm i\epsilon^{IJ}_{KL}\right)$$ carry
out the following homomorphism:
\be
   \mathfrak{so}(1,3) = \mathfrak{sl}(2,C)_R \oplus
   \mathfrak{sl}(2,C)_L.
                                 \lb{15} \ee
As a result of Eq.~(\ref{15}), non-zero components of connections
are only $A^{(\pm) i } = A^ {(\pm) 0i}$. Instead of
(anti-)self-duality, the terms of left-handed (+) and right-handed
(-) components are used.

Plebanski \ct{3} and other authors \ct{3a,3b} suggested to
consider a gravitational action in the (visible) world as a
left-handed $\mathfrak{sl}(2,C)_L^{(grav)}$- invariant action,
which contains self-dual fields $F=F^{(+)i}$ and
$\Sigma=\Sigma^{(+)i}$ (i=1,2,3):
\be I_{(grav)}(\Sigma,A,\psi) = \frac{1}{\kappa^2} \int
\left[\Sigma^i\wedge F^i +
 \left(\Psi^{-1}\right)_{ij}\Sigma^i\wedge \Sigma^j\right].
                      \lb{18} \ee
Here $\Sigma^i=2B^{0i}$, and $\Psi_{ij}$ are auxiliary fields,
defining a gauge, which provides equivalence of Eq.~(\ref{18}) to
the Einstein-Hilbert gravitational action:
\be I_{(EG)} = \frac{1}{\kappa^2} \int d^4 x (\frac{R}{2} -
\Lambda ),
                           \lb{19} \ee
where $R$ is a scalar curvature, and $\Lambda$ is the Einstein
cosmological constant.

\section{Graviweak unification model}

On a way of unification of the gravitational and weak interactions
we considered an extended $\mathfrak g = \mathfrak
{spin}(4,4)$-invariant Plebanski's action:
\be I(A, B, \Phi) = \frac{1}{g_{uni}} \int_{\mathfrak
M}\left\langle BF +  B\Phi B + \frac 13 B\Phi \Phi \Phi B
\right\rangle, \lb{22} \ee
where $\langle...\rangle$ means a wedge product, $g_{uni}$ is an
unification parameter, and $\Phi_{IJKL}$ are auxiliary fields.

Varying the fields $A,B$ and $\Phi$, we obtained the field
equations:
\be  {\cal D}B = dB + [A,B] = 0,  \lb{23} \ee
where ${\cal D}$ is the covariant derivative, $${\cal
D}_{\mu}^{IJ} = \delta^{IJ}\partial_{\mu} - A_{\mu}^{IJ},$$ and
\be   F = -2\left(\Phi +\frac 13\Phi \Phi \Phi\right)B,  \lb{24}
\ee
\be    B^{IJ} B^{KL} = - \frac 1{16} B^{IJ}
\Phi^{KL}_{MN}\Phi^{MN}_{PQ}B^{PQ}. \lb{25} \ee
The first equation describes the dynamics, while last two
determine $B$ and $\Phi$ respectively. Here we assumed that the
cosmological constant is zero: $\Lambda=0$.

Having considered the equations of motion, obtained by means of
the action (\ref {22}), and having chosen a possible class of
solutions, we can present the following action for the Graviweak
unification (see details in Refs.~\ct{1,2}):
\be   I(A, \Phi) = \frac{1}{8g_{uni}} \int_{\mathfrak M} \langle
\Phi FF \rangle, \lb{38} \ee
where
\be  \langle \Phi F F \rangle =
\frac{d^4x}{32}\epsilon^{\mu\nu\rho\sigma}{{\Phi_{\mu\nu}}^{\varphi\chi
IJ}}_{KL}F_{\varphi\chi IJ} {F_{\rho\sigma}}^{KL},  \lb{39} \ee
and
\be {{\Phi_{\mu\nu}}^{\rho\sigma ab}}_{cd} = (e_{\mu}^f)
(e_{\nu}^g){\epsilon_{fg}}^{kl}(e^{\rho}_k)
(e^{\sigma}_l)\delta_{cd}^{ab}. \lb{41} \ee
A spontaneous symmetry breaking of our new action that produces
the dynamics of gravity, weak $SU(2)$ gauge and Higgs fields,
leads to the conservation of the following sub-algebra:
$$  \tilde {\mathfrak g} = {{\mathfrak sl}(2,C)}^{(grav)}_L
\oplus {\mathfrak su}(2)_L.$$ Considering indices $a, b \in
\{0,1,2,3\}$ as corresponding to $I,J=1,2,3,4$, and indices $m, n$
as corresponding to indices $I, J=5,6,7,8$, we can present a
spontaneous violation of the Graviweak unification symmetry in
terms of the 2-forms: $$A = \frac 12 \omega + \frac 14 E + A_W,$$
where $\omega = \omega^{ab}\gamma_{ab}$ is a gravitational
spin-connection, which corresponds to the sub-algebra $\mathfrak
{sl}(2,C)_L^{(grav)}$. The connection $E = E^{am}\gamma_{am}$
corresponds to the non-diagonal components of the matrix $A^{IJ}$,
described by the following way (see \ct{2}): $E = e\varphi =
e^a_{\mu}\gamma_a\varphi^m\gamma_m dx^{\mu}$. The connection $A_W
= \frac 12 A^{mn}\gamma_{mn}$ gives: $A_W = \frac 12 A_W^i
\tau_i$, which corresponds to the sub-algebra $\mathfrak
{su}(2)_L$ of the weak interaction. Here $\tau_i$ are the Pauli
matrices with $i=1,2,3$.

Assuming that we have only scalar field
$\varphi^m=(\varphi,\varphi^i)$, we can consider a symmetry
breakdown of the Graviweakk Unification, leading to the following
OW-action \ct{1}:
$$I_{(OW)}\left(e,\varphi,A,A_W\right)= \frac{3}{8g_{uni}}
\int_{\bf M} d^4x|e|\left(\frac 1{16}{|\varphi|}^2 R -
\frac{3}{32}{|\varphi|}^4 \right.$$
 \be + \frac 1{16}{R_{ab}}^{cd}
{R^{ab}}_{cd} - \left.\frac 12 {\cal D}_a{\varphi^{\dagger}} {\cal
D}^a\varphi - \frac 14 {F_W^i}_{ab}{F_W^i}^{ab} \right). \lb{27u}
\ee
In Eq.~(\ref{27u}) we have the Riemann scalar curvature $R$;
$|\varphi |^2 = {\varphi }^{\dag}\varphi$ is a squared scalar
field, which from the beginning is not the Higgs field of the
Standard Model; ${\cal D }\varphi = d\varphi + [A_W, \varphi]$ is
a covariant derivative of the scalar field, and $F_W = dA_W +
[A_W, A_W]$ is a curvature of the gauge field $A_W$. The third
term of the action (\ref{27u}) belongs to the Gauss-Bone theory of
gravity (see for example Refs.~ \ct{4,4a}).

Eq.~(\ref{27u}) allows us to return to the GR formalism, when the
dynamics is described by the metric tensor $g_{\mu\nu}$, and we
have:
$$I_{(OW)}\left(\varphi,A,A_W\right)= \frac{3}{8g_{uni}}
\int_{\bf M} d^4x\sqrt{-g}\left(\frac 1{16}{|\varphi|}^2 R -
\frac{3}{32}{|\varphi|}^4 \right.$$
 \be + \frac 1{16}R_{\alpha\beta\mu\nu}
R^{\alpha\beta\mu\nu} - \left.\frac 12 {\cal
D}_{\mu}{\varphi^{\dagger}} {\cal D}^{\mu}\varphi - \frac 14
{F_W^i}_{\mu\nu}{F_W^i}^{\mu\nu} \right). \lb{27v} \ee
Here we consider the Jordan frame, in accordance with Ref.~\ct{2}.

\subsection{Existence of de Sitter solutions at the early time of the Universe}

It is well-known that the early time acceleration era of the
Universe is described by (quasi)-de Sitter solutions (see for
example \ct{Chen,Odin}). So firstly it is important to investigate
if de Sitter solutions exist in the case of the action
(\ref{27u}). Such a problem was investigated by authors of
Ref.~\ct{2}. Taking into account that our model \ct{1} is a
special case of the more general $SU(N)$ model \ct{2} of the
unification of gravity, gauge fields, and Higgs bosons, we can
assume that the Universe is inherently de Sitter, where the
4-spacetime is a hyperboloid in a 5-dimensional Minkowski space
under the constraint
\be - x^2_0 + x^2_1 + x^2_2 + x^2_3 + x^2_4 = r^2_{dS}, \lb{10dS}
\ee
where $r_{dS}$ is the radius of curvature of the de Sitter space,
or simply the de Sitter radius. The Hubble expansion of the
Universe is then viewed as a process that approaches the
asymptotic limit of a pure space which is de Sitter in nature,
evidenced that the cosmological constant (CC) describes the dark
energy (DE) substance, which has become dominant in the Universe
at late times:
\be  \Omega_{DE} = \rho_{DE}/\rho_{crit} \simeq 0.75,  \lb{11dS}
\ee
where $\rho$ is the energy density and the critical density is
\be  \rho_{crit} = \frac{3H_0^2}{8\pi G_N}\simeq 1.88 \times
10^{-29} H_0^2, \lb{12dS} \ee
where $H_0$ is the Hubble constant:
\be H_0 \simeq 1.5\times 10^{-42}\,\,{\rm{GeV}}. \lb{13dS} \ee
Identifying the Einstein tensor as
\be G_{\mu\nu} = - \frac{3}{r^2_{dS}}g_{\mu\nu}, \lb{14dS} \ee
we see that the only nontrivial component that satisfies this
equation is a constant for the Ricci scalar:
\be R_0=\frac{12}{r^2_{dS}}, \lb{15dS} \ee
and
\be R_{\alpha\beta\mu\nu}R^{\alpha\beta\mu\nu} =\frac 16 R_0^2
=\frac{24}{r^4_{dS}}. \lb{16dS} \ee
As it was shown in Ref.~\ct{2}, the nontrivial vacuum solution to
the action (\ref{27u}) is de Sitter spacetime with a non-vanishing
Higgs vacuum expectation value (VEV) of the scalar field:
$v=\langle\varphi\rangle =\varphi_0$. The standard Higgs potential
in Eq.~(\ref{27u}) has an extremum at $\varphi_0^2=R_0/3$ (with
$R_0
> 0$), corresponding to a de Sitter spacetime background solution:
\be R_0=\frac{12}{r^2_{dS}} = 3v^2 = 4\Lambda_0, \lb{17dS} \ee
which implies vanishing curvature:
\be F_0 = \frac 12 R_0 - \frac 1{16}\Sigma_0\varphi_0^2, \lb{18dS}
\ee
solving the field equations ${\cal D}F = dF + [A,F] = 0$, and
strictly minimizing the action (\ref{27u}).

Based on this picture, the origin of the cosmological constant
(and also DE) is associated with the inherent spacetime geometry,
and not with vacuum energy of particles (we consider their
contributions later). Note that as a fundamental constant under de
Sitter symmetry, $r_{dS}$ is not subject to quantum corrections.

Local dynamics then exist as fluctuations with respect to this
cosmological background. In general, de Sitter space may be
inherently unstable. The quantum instability of de Sitter space
was investigated by various authors.  Abbott and Deser \ct{AD}
have shown that de Sitter space is stable under a restricted class
of classical gravitational perturbations. So any instability of de
Sitter space may likely have a quantum origin. Ref.~\ct{Ford}
demonstrated through the expectation value of the energy-momentum
tensor for a system with a quantum field in a de Sitter background
space that in general it contains a term that is proportional to
the metric tensor and grows in time. As a result, the curvature of
the spacetime would decrease and de Sitter space tends to decay
into the flat space (see similar conclusions in Ref.~\ct{AIT}). We
note that the expectation value of such energy-momentum tensor is
\be  \langle T_{\mu\nu}\rangle \propto g_{\mu\nu}H^4(Ht),
                                                   \lb{19dS} \ee
where H is the Hubble parameter in the de Sitter metric, and
\be  ds^2 = dt^2 - a^2(t)d{\bf x}^2            \lb{20dS} \ee
with $a(t) = e^{Ht}$. According to (\ref{19dS}), the decay time of
this process is
\be   \tau \sim H^{-1}. \lb{21dS} \ee
In our case, this means that the decay time is of the order of the
de Sitter radius:
\be   \tau \sim r_{dS} \simeq 1.33H_0^{-1}. \lb{22dS} \ee
Since the age of our universe is smaller than $r_{dS}$, we are
still observing the accelerating expansion in action.

Of course, we can consider the perturbation solutions of the de
Sitter solution:
\be   H(t) = H_{dS} + \Delta H(t), \lb{23dS} \ee
where the perturbation is very small: $|\Delta H(t)| << 1$. The
evolution of linear perturbation can behave as
\be  \Delta H(t) = c + c_1 e^{\delta_1 t} + c_2 e^{\delta_2 t} ,
                                       \lb{24dS} \ee
(see for example Ref.~\cite{Odin}). But we don't concern this
problem in this paper.

\subsection{Parameters of Graviweak unification model}

According to (\ref{27u}), the Newton gravitational constant $G_N$
is defined by the expression:
\be 8\pi G_N = ({M^{(red.)}_{Pl}})^{-2} = \frac{64g_{uni}}{3v^2},
\lb{28u} \ee
a bare cosmological constant is equal to
\be \Lambda_0 = \frac 34 v^2, \lb{29u} \ee
and
\be g_W^2 = 8g_{uni}/3. \lb{30u} \ee
The coupling constant $g_W$ is a bare coupling constant of the
weak interaction, which also coincides with a value of the
constant $g_2=g_W$ at the Planck scale. Considering the running
$\alpha_2^ {-1}(\mu)$, where $\alpha_2=g_2^2/4\pi$, we can carry
out an extrapolation of this rate to the Planck scale, what leads
to the following estimation \ct{5,5a}:
\be \alpha_2 (M_{Pl}) \sim 1/50,  \lb{31u} \ee
and then the overall GWU parameter is: $g_{uni}\sim 0.1.$

According to Eqs.~(\ref{27v}) and (\ref{28u}), we obtain the
following GWU action:
$$I_{(OW)}\left(\varphi,A,A_W\right)=\int_{\bf M}
d^4x\sqrt{-g}\big[
\left(\frac{M^{red}_{Pl}}{v}\right)^2\left(\frac 12{|\varphi|}^2 R
- \frac 34{|\varphi|}^4
 + \frac 12R_{\alpha\beta\mu\nu}
R^{\alpha\beta\mu\nu}\right)$$ \be - \frac{1}{g_W^2}\left (\frac
12 {\cal D}_{\mu}{\varphi^{\dagger}} {\cal D}^{\mu}\varphi + \frac
14 {F_W^i}_{\mu\nu}{F_W^i}^{\mu\nu} \right)\big]. \lb{32v} \ee
In the action (\ref{32v}) the Lagrangian includes the non-minimal
coupling with gravity
\ct{1Bez,2Bez,3Bez,4Bez,5Bez,6Bez,1B,2B,3B,4B,5B,FGH}. We see that
the field $\varphi$ is not stuck at $\varphi_0$ anymore, but it
can be represented as
\be \varphi = \varphi_0 - \sigma = v - \sigma ,  \lb{1z} \ee
where the real scalar field $\sigma$ is an {\un {inflaton}}. Here
we see that in the minimum, when $\varphi=v$, the inflaton field
is zero ($\sigma=0$), and then increases with the falling of the
field $\varphi$.

Inserting designations (\ref{1z}) into the action (\ref{32v}), we
obtain:
$$I_{(OW)}\left(\varphi,A,A_W\right)=\int_{\bf M}
d^4x\sqrt{-g}\big[
\left(\frac{M^{red}_{Pl}}{v}\right)^2\left(\frac 12(v - \sigma)^2
R - \frac 34(v - \sigma)^4
 + \frac 12R_{\alpha\beta\mu\nu}
R^{\alpha\beta\mu\nu}\right)$$ \be - \frac{1}{g_W^2}\left (\frac
12 {\cal D}_{\mu}(v - \sigma) {\cal D}^{\mu}(v - \sigma) + \frac
14 {F_W^i}_{\mu\nu}{F_W^i}^{\mu\nu} \right)\big], \lb{2z} \ee
or (in a suitable gauge condition):
$$I_{(OW)}\left(\varphi,A,A_W\right)=\int_{\bf M}
d^4x\sqrt{-g}\big[\left(M_{Pl}^{red}\right)^2 \left(\frac 12(1 -
\frac{\sigma}{v})^2 R - \Lambda_0(1 - \frac{\sigma}{v})^4
 + \frac 1{2v^2}R_{\alpha\beta\mu\nu}
R^{\alpha\beta\mu\nu}\right)$$ \be - \frac{1}{g_W^2}\left (\frac
12 {\cal D}_{\mu}{\sigma} {\cal D}^{\mu} \sigma + \frac 14
{F_W^i}_{\mu\nu}{F_W^i}^{\mu\nu} \right)\big], \lb{3z} \ee
Considering the expansion in powers of small values of $\sigma/v$,
and leaving only the first-power terms, we can present the
following gravitational part of the action near the Planck scale:
$$I_{(grav\,\, OW)} = \int_{M}d^4x
\sqrt{-g}\big[\left(M_{Pl}^{red}\right)^2 \left( \frac 12 R -
\Lambda_0 - (R - 4\Lambda_0)\frac{\sigma}{v} + (\frac{R}{2} -
6\Lambda_0)\frac{\sigma^2}{v^2} + ...\right)$$ \be -
\frac{1}{2g_W^2}{\cal D}_{\mu}{\sigma} {\cal D}^{\mu} \sigma +
...\big]. \lb{4z} \ee
As it was shown in Subsection 3.1, near the minimum at the Planck
scale we have: $$R=R_0 + \Delta R,$$ where $\Delta R << R_0$. Here
$\Lambda_0 = \frac 34 v^2 = R_0/4$. Using the last relations, we
can neglect the third term in Eq.~(\ref{4z}), i.e $\Delta R
\sigma/v$, and obtain:
$$ I_{(grav\,\, OW)} = \int_{M}d^4x
\sqrt{-g}\left(M_{Pl}^{red}\right)^2 \left(\frac 12 (1 +
\frac{\sigma^2}{v^2})R - \Lambda_0 -
\frac{6\Lambda_0}{v^2}\sigma^2 + ...\right)$$ \be -
\frac{1}{2g_W^2}{\cal D}_{\mu}{\sigma} {\cal D}^{\mu} \sigma +
...\big].  \lb{5z} \ee
Now it is possible to get rid of the non-minimal coupling to
gravity  by making the conformal transformation from the Jordan
frame of Eq.~(\ref{5z}) to the Einstein frame (see
Refs.~\ct{1Bez,5B}).

According to Refs.~\ct{1Bez,5B}, in units $M_{Pl}^{red}=1$, the
Lagrangian including the generalized non-minimal coupling to
gravity reads:
\be L_J = \sqrt{-g}\left(\frac 12 \Omega(\chi)R - \Lambda_0 -
\frac 12 (\partial \chi)^2 - V_J(\chi) + ...\right),
         \lb{6z} \ee
where
\be \Omega = 1 + \xi \chi^2.  \lb{7z} \ee
In our case:
\be \chi = \frac{\sigma}{g_W}, \lb{71z} \ee
and
\be \xi = \frac {g_W^2}{v^2} = \frac 18, \lb{72z} \ee
and
\be V_J(\chi) = \frac 12 m^2 \chi^2, \lb{73z} \ee
where
\be m^2=\frac 98 v^2 \lb{74z} \ee
is a bare mass of the inflaton.

In order to transform from the Jordan frame to the canonical
Einstein frame, we need to redefine the metric:
\be  \hat{g}_{\mu\nu} = \Omega g_{\mu\nu}.  \lb{8z} \ee
Finally, the Larangian in the Einstein frame has the form:
\be L_E = \sqrt{-\hat g}\left(\frac 12 \hat R - \Lambda_0 - \frac
12 (\partial \chi)^2 - V_E(\chi) + ...\right),
         \lb{9z} \ee
where $$ V_E(\chi) = \frac{V_J(\chi)}{\Omega^2(\chi)}.$$ In our
case we have small values of the field $\chi$, $\Omega \simeq 1$
(see \ct{1Bez}), and in the Einstein frame our action (\ref{5z})
near the Planck scale minimum is
$$ I^{(E)}_{(grav\,\, OW)} \simeq \int_{M}d^4x
\sqrt{-g}\big[\left(M_{Pl}^{red}\right)^2 \left(\frac 12 R -
\Lambda_0 - \frac 12 m^2 \chi^2 + \frac
1{2v^2}R_{\alpha\beta\mu\nu} R^{\alpha\beta\mu\nu}\right)$$ \be -
\frac 12{\cal D}_{\mu}\chi {\cal D}^{\mu}\chi - \frac 1{4g_W^2}
{F_W^i}_{\mu\nu}{F_W^i}^{\mu\nu} \big]. \lb{10z} \ee

\section{Multiple Point Model}

The radiative corrections to the effective Higgs potential,
considered in Refs.~\ct{6,7}, bring to the emergence of the second
minimum of the effective Higgs potential at the Planck scale. It
was shown that in the 2-loop approximation of the effective Higgs
potential, experimental values of all running coupling constants
in the SM predict an existence of the second minimum of this
potential located near the Planck scale, at the value $v_2
=\varphi_{min2}\sim M_{Pl}$.

In general, a quantum field theory allows an existence of several
minima of the effective potential, which is a function of a scalar
field. If all vacua, corresponding to these minima, are
degenerate, having zero cosmological constants, then we can speak
about the existence of a multiple critical point (MCP) at the
phase diagram of theory considered for the investigation (see
Refs.~\ct{8,8a}).  In Ref.~\ct{8} Bennett and Nielsen suggested
the Multiple Point Model (MPM) of the Universe, which contains
simply the SM itself up to the scale $\sim 10^{18}$ GeV. In
Ref.~\ct{9} the MPM was applied (by the consideration of the two
degenerate vacua in the SM) for the prediction of the top-quark
and Higgs boson masses, which gave:
\be M_t = 173 \pm 5 \,\, {\rm GeV }, \qquad M_H = 135 \pm 9 \,\,
{\rm GeV }. \lb{29} \ee
Later, the prediction for the mass of the Higgs boson was improved
by the calculation of the two-loop radiative corrections to the
effective Higgs potential \ct{6,7}. The predictions: 125 GeV
$\lesssim M_H \lesssim$ 143 GeV in Ref.~\ct {6}, and 129 $\pm$ 2
GeV in Ref.~\ct{7} -- provided the possibility of the theoretical
explanation of the value $M_H\approx$ 126 GeV observed at the LHC.
The authors of Ref.~\ct{7a} have shown that the most interesting
aspect of the measured value of $M_H$ is its near-criticality.
They have thoroughly studied the condition of near-criticality in
terms of the SM parameters at the high (Planck) scale. They
extrapolated the SM parameters up to large energies with full
3-loop NNLO RGE precision. All these results mean that the
radiative corrections to the Higgs effective potential lead to the
value of the Higgs mass existing in the Nature.

The behavior of the Higgs self-coupling $\lambda$ is quite
peculiar: it decreases with energy to eventually arrive to a
minimum at the Planck scale values and then starts to increase
there after. Within the experimental and theoretical uncertainties
the Higgs coupling $\lambda$ may stay positive all way up till the
Planck scale, but it may also cross zero at some scale $\mu_0$. If
that happens, our Universe becomes unstable.

The largest uncertainty in couplings comes from the determination
of the top Yukawa coupling. Smaller uncertainties are associated
to the determination of the Higgs boson mass and the QCD coupling
$\alpha_s$ (see Refs.~\ct{6,7,7a}).\\
Calculations of the lifetime of the SM vacuum are extremely
sensitive to the Planck scale physics. The authors of
Refs.\ct{1Br,2Br,3Br} showed that if the SM is valid up to the
Planck scale, then the Higgs potential becomes unstable at $\sim
10^{11}$ GeV. There are two reasons of this instability. In
typical tunnelling calculations, the value of the field at the
center of the critical bubble is much larger than the point of the
instability. In the SM case, this turns out to be numerically
within an order of magnitude of the Planck scale.

The measurements of the Higgs mass and top Yukawa coupling
indicate that we live in a very special Universe: at the edge of
the absolute stability of the EW vacuum. If fully stable, the SM
can be extended all the way up to the inflationary scale and the
Higgs field, non-minimally coupled to gravity with strength $\xi$,
can be responsible for the inflation (see Ref.~\ct{1Bez}).

Having substituted in Eq.(\ref{28u}) the values of $g_ {uni}
\simeq 0.1$ and $G_N=1/8\pi {(M_{Pl}^{red.})}^2$, where $M_{Pl}^{
red.}\approx 2.43\cdot 10^{18}$ GeV, it is easy to obtain the
VEV's value $v$, which in this case is located near the Planck
scale:
\be v=v_2\approx 3.5\cdot 10^{18} {\rm{GeV}}. \lb{32u} \ee
Such a result takes place, if the Universe at the early stage
stayed in the "false vacuum", in which the VEV of the Higgs field
is huge: $v=v_2\sim 10^{18} GeV$. The exit from this state could
be carried out only by means of the existence of the second scalar
field. In the present paper we assume that the second scalar
field, participating into the Inflation, is the mirror Higgs
field, which arises from the interaction between the Higgs fields
of the visible and invisible sectors of the Universe.

\section{Mirror world with broken mirror parity}

As it was noted at the beginning of this paper, we assumed the
parallel existence in the Nature of the visible (OW) and invisible
(MW) (mirror) worlds.

Such a hypothesis was suggested in Refs.~\ct{LY,KOP}.

The Mirror World (MW) is a mirror copy of the Ordinary World (OW)
and contains the same particles and types of interactions as our
visible world, but with the opposite chirality. Lee and Yang
\ct{LY} were first to suggest such a duplication of the worlds,
which restores the left-right symmetry of the Nature. The term
``Mirror Matter'' was introduced by Kobzarev, Okun and Pomeranchuk
\ct{KOP}, who first suggested to consider MW as a hidden
(invisible) sector of the Universe, which interacts with the
ordinary (visible) world only via gravity, or another (presumably
scalar) very weak interaction.

In the present paper we consider the hidden sector of the Universe
as a Mirror World (MW) with broken Mirror Parity (MP)
\ct{1mw,2mw,3mw,4mw,5mw}. If the ordinary and mirror worlds are
identical, then O- and M-particles should have the same
cosmological densities. But this is immediately in conflict with
recent astrophysical measurements \ct{1D,2D,3D}. Astrophysical and
cosmological observations have revealed the existence of the Dark
Matter (DM), which constitutes about 25\% of the total energy
density of the Universe. This is five times larger than all the
visible matter, $\Omega_{DM}: \Omega_{M} \simeq 5 : 1$. Mirror
particles have been suggested as candidates for the inferred dark
matter in the Universe \ct{4mw,6mw,7mw} (see also \ct{Sil}).
Therefore, the mirror parity (MP) is not conserved, and the OW and
MW are not identical.

The group of symmetry $G_ {SM}$ of the Standard Model was enlarged
to $G_ {SM}\times G'_ {SM'}$, where $G_{SM}$ stands for the
observable SM, while $ G'_ {SM'}$ is its mirror gauge counterpart.
Here O(M)- particles are singlets of the group $G'_ {SM'}$ ($G _
{SM}$).

It was assumed that the VEVs of the Higgs doublets $\phi$ and
$\phi'$ are not equal \ct{1mw,2mw,3mw,4mw,5mw}:
$$\langle\phi\rangle=v, \quad \langle\phi' \rangle=v', \quad {\rm
{and} }\quad v\neq v'. $$ The parameter characterizing the
violation of the MP is $\zeta = {v'}/{v} \gg 1$. Astrophysical
estimates give: $\zeta > 30, \quad \zeta \sim 100$ (see
Refs.~\ct{10mw,11mw} and references there).

The action $I_{(MW)}$ in the mirror world is represented by the
same integral (\ref{27u}), in which we have to make the
replacement of all OW-fields by their mirror counterparts: $e,
\varphi, A, A_W, R \to e', \varphi', A', A' _W, R'$.

In general, ordinary and mirror matters interact with gravity by
two separate metric tensors $g^L_{\mu\nu}$ and $g^R_{\mu\nu}$,
i.e. each sector of the Universe has its own GR-like gravity.

In this investigation we assume (see \ct{BLNT} and \ct{LNT}) that
the left-handed gravity coincides with the right-handed gravity:
the left-handed and right-handed connections are equal: $A'= A$,
i.e. $A^L = A^R$. In this case $g^L_{\mu\nu}=g^R_{\mu\nu}$, and
the left-handed and right-handed gravity equally interact with
visible and mirror matters.

Then in the Einstein frame, the MW-action near the Planck scale
minimum is
$$ I^{(E)}_{(grav\,\, MW)} \simeq \int_{M}d^4x
\sqrt{-g}\big[\left(M_{Pl}^{red}\right)^2 \left(\frac 12 R' -
\Lambda'_0 - \frac 12 {m'}^2 {\chi'}^2 + \frac
1{2{v'}^2}R'_{\alpha\beta\mu\nu} {R'}^{\alpha\beta\mu\nu}\right)$$
\be - \frac 12{\cal D}_{\mu}\chi' {\cal D}^{\mu}\chi' - \frac
1{4{g'}_W^2} {{F'}_W^i}_{\mu\nu}{{F'}_W^i}^{\mu\nu} \big].
\lb{10'z} \ee
>From Eq.~(\ref{10'z}), it is not difficult to determine that in
the hidden (mirror) world the Newton gravitational constant $G'_N$
is defined by the expression:
\be 8\pi G'_N = ({M'}^{(red.)}_{Pl})^{-2} =
\frac{64g'_{uni}}{3v'^2}, \lb{28u'} \ee
a bare cosmological constant is equal to
\be \Lambda'_0 = \frac 34 v'^2, \lb{29u'} \ee
and
\be {g'}_W^2 = 8g'_{uni}/3. \lb{30u'} \ee
If we assume the same gravity in the OW and MW, then we have:
\be G'_N=G_N,\quad {M'}^{(red.)}_{Pl} = M^{(red.)}_{Pl}, \lb{31u'}
\ee
what means that
\be \frac{g'_{uni}}{v'^2} = \frac {g_{uni}}{v^2}, \lb{32u'} \ee
\be    g'_{uni}\neq g_{uni},\quad  g'_W\neq g_W.
 \lb{33u'} \ee
Then:
\be {g'}_W^2 = \frac{v'^2}{v^2}g_W^2 = \zeta^2 g_W^2 , \lb{34u'}
\ee
and
\be \Lambda'_0 = \zeta^2\Lambda_0. \lb{35u'} \ee
It is well-known that the hidden (invisible) sector of the
Universe interacts with the ordinary (visible) world only via
gravity, or another very weak interaction (see for example
\ct{KOP,KST,7mw}). In particular, the authors of Ref.~\ct{12mw}
assumed, that along with gravitational interaction there also
exists the interaction between the initial Higgs fields of both
OW- and MW-worlds:
\be V_{int} = \alpha_\chi
(\chi^{\dagger}\chi)({\chi'}^{\dagger}\chi'). \lb{30} \ee
Taking into account the interaction (\ref{30}) and OW- and MW-
actions (\ref{10z}), (\ref{10'z}), we obtain the total action of
the Universe (in the Einstein frame):
$$ I^{(E)}_{(tot)} \simeq \int_{M}d^4x
\sqrt{-g}\big[\left(M_{Pl}^{red}\right)^2 \left(\frac 12 (R+R') -
(\Lambda_0+ \Lambda'_0) - \frac 12 (m^2 \chi^2 + {m'}^2 {\chi'}^2)
- \alpha_\chi (\chi^2)({\chi'}^2) \right)$$
$$  + \frac 1{16}\left(\frac 1{ g_W^2}R_{\alpha\beta\mu\nu}
R^{\alpha\beta\mu\nu} + \frac 1{ {g'}_W^2}R'_{\alpha\beta\mu\nu}
{R'}^{\alpha\beta\mu\nu}\right)$$ \be - \frac 12{\cal D}_{\mu}\chi
{\cal D}^{\mu}\chi - \frac 12{\cal D'}_{\mu}\chi' {\cal
D'}^{\mu}\chi' - \frac 1{4g_W^2} {F_W^i}_{\mu\nu}{F_W^i}^{\mu\nu}
- \frac 1{4{g'}_W^2} {{F'}_W^i}_{\mu\nu}{{F'}_W^i}^{\mu\nu} +
L_{SM} + L'_{SM'}\big], \lb{31} \ee
where $L_{SM}$ and $L'_{SM'}$ are the $SM$ and $SM'$ matter
Lagrangians, respectively.

\section{Cosmological constant problem}

In the Einstein-Hilbert action the vacuum energy is:
\be \rho_{vac} = \left(M_{Pl}^{red}\right)^2 \Lambda, \lb{7y} \ee
where $\Lambda$ is the cosmological constant of the Universe.

In our case, given by Eq.~(\ref{31}), the vacuum energy density is

 \be \rho_0 =
 \left(M_{Pl}^{red}\right)^2 (\Lambda_0 + \Lambda'_0) =  \left(M_{Pl}^{red}\right)^2
 (1 + \zeta^2)\Lambda_0. \lb{8y} \ee
However, assuming the existence of the discrete spacetime of the
Universe at the Planck scale and using the prediction of the
non-commutativity suggested  by B.G. Sidharth \ct{1S,2S}, we
obtain that the gravitational part of the GWU action has the
vacuum energy density equal to zero, or almost zero.

Indeed, the total cosmological constant and the total vacuum
density of the Universe contain also the vacuum fluctuations of
fermions and other SM boson fields:
\be \Lambda \equiv \Lambda_{eff} = \Lambda^{ZMD} + (\Lambda_0 +
\Lambda'_0) - \Lambda^{(NC)}_s + \Lambda^{(NC)}_f, \lb{9y} \ee
where $\Lambda^{ZMD}$ is zero mode degrees of freedom of all
fields existing in the Universe, and $\Lambda^{(NC)}_{s,f}$ are
boson and fermion contributions of the non-commutative geometry of
the discrete spacetime at the Planck scale. If according to the
theory by B.G. Sidharth \ct{1S,2S}, we have:
\be \rho_{vac}^{(0)} = \left(M_{Pl}^{red}\right)^2 \Lambda^{(0)} =
 \left(M_{Pl}^{red}\right)^2 \left(\Lambda^{ZMD} +
(\Lambda_0 + \Lambda'_0)- \Lambda_s^{(NC)}\right) \approx 0,
\lb{10y} \ee
then Eq.~(\ref{31}) contains the cosmological constant
$\Lambda^{(0)} \approx 0$. In Eqs.~(\ref{9y}) and (\ref{10y}) the
bosonic (scalar) contribution of the non-commutativity is:
\be \rho_{(scalar)}^{(NC)} \simeq m_s^4 \quad ({\rm{in \,
units}}:\, \hbar = c = 1), \lb{11y} \ee
which is given by the mass $m_s$ of the primordial scalar field
$\varphi$. Then the discrete spacetime at the very small distances
is a lattice (or has a lattice-like structure) with a parameter $a
= \lambda_s = 1/m_s.$ This is a scalar length:
$$
a = \lambda_s \sim 10^{-19}\,\,{\rm GeV}^{-1},$$ which coincides
with the Planck length: $\lambda_{Pl} = 1/M_{Pl} \approx
10^{-19}\,\,{\rm GeV}^{-1}.$
The assumption:
\be \Lambda^{(0)} = \Lambda^{ZMD} + (\Lambda_0 + \Lambda'_0) -
\Lambda_s^{(NC)} \approx 0 \lb{12y} \ee
means that the Graviweak Unification model contains the
cosmological constant equal to zero, or almost zero.

B.G.~Sidharth gave in Ref.~\ct{3S} the estimation:
\be \rho_{DE} = \left(M_{Pl}^{red}\right)^2 \Lambda_f^{(NC)},
\lb{13z} \ee
considering the non-commutative contribution of light primordial
neutrinos as a dominant contribution to $\rho_{DE}$, which
coincides with astrophysical measurements \ct{1D,2D,3D}:
\be \rho_{DE} \approx (2.3\times 10^{-3}\,\,\rm{eV})^4. \lb{14y}
\ee
Returning to the inflation model, we rewrite the action (\ref{31})
as:
\be I_{(tot)} = \int_{\large M} d^4x \sqrt{-g}
\left[\left(M_{Pl}^{red}\right)^2 \left(\frac 12 (R+R') - \Lambda
-\frac {m^2}{2} \chi^2 - \frac{{m'}^2}{2} {\chi'}^2 - \alpha_\chi
(\chi^2)({\chi'}^2) \right) + ... \right], \lb{15y} \ee
where the positive cosmological constant $\Lambda$ is not zero,
but is very small.

\section{Inflation from the Higgs Field False Vacuum}

Inflation from a local minimum is a viable scenario of the hybrid
model by A.~Linde \ct{Lin} provided a graceful exit to the
radiation-dominated era, which can be obtained via some mechanism
beyond the SM. It was shown in Refs.~\ct{LNS} and \ct{3FV,4FV}
that this mechanism is consistent only with a narrow range of
values of the Higgs boson mass, indicated experimentally by ATLAS
and CMS \ct{1Hig,2Hig}, and also depends on a narrow range of the
top-quark mass values.

 The existence of the second Higgs field $\chi'$
could be the cause of the hybrid inflation (see \ct{Lin}),
bringing the Universe out of the ``false vacuum" with the VEV
$v_2\sim 10^{18}$ GeV. This circumstance provided the subsequent
transition to the vacuum with the Higgs VEV $v_1$ existing at the
Electroweak (EW) scale. Here $v_1\approx 246$ GeV is a vacuum, in
which we live at the present time.

Using the action (\ref{15y}), given by GWU, we obtain near the
local ``false vacuum'' the following gravitational potential in
units $M_{Pl}^{red}=1$:
\be   V(\chi, \chi') = \Lambda + \frac {m^2}{2} \chi^2 +
\frac{{m'}^2}{2} {\chi'}^2 + \alpha_\chi (\chi^2)({\chi'}^2).
\lb{16y} \ee
The local minimum of the potential (\ref{16y}) at $\varphi=v$
(when $\chi=0$) and ${\varphi'}_0 \neq v'$ ($\chi' \neq 0$),
gives:
 \be  V( 0,\,\, \chi')= \Lambda + \frac{{m'}^2}2(\chi')^2.
                           \lb{31z} \ee
The last equation (\ref{31z}) shows that the potential
$V(0,\chi')$ grows with growth of $\chi'$, i.e. with falling of
the field $\varphi'$. It means that a barrier of potential grows
and at some value $\chi' = \chi'_{in}$ the Higgs field $\varphi'$
begins its inflationary falling.

The local minimum at $\chi=0$ and $\chi'=\chi'|_{in}$ is given by
the following condition:
\be \frac{\partial V(\chi, \chi')}{\partial \chi^2}\big|_{\chi=0}
= \frac{m^2}{2} + \alpha_\chi {\chi'}_{in}^2 = 0,
    \lb{22z} \ee
and
 \be \frac{\partial V(\chi, \chi')}{\partial{\chi'}^2}\big|_ {\chi'=0} =
\frac{{m'}^2}{2} + \alpha_\chi {\chi}_{in}^2 = 0,
    \lb{23y} \ee
what gives:
 \be {\chi}_{in}^2 = - \frac {{m'}^2}{2\alpha_\chi} \quad {\rm{and}}
 \quad {\chi'}_{in}^2 = - \frac {{m}^2}{2\alpha_\chi}.    \lb{23a} \ee
If $\alpha_\chi < 0$, we have the following value of the barrier
of potential, at which the inflationary falling of the Higgs field
$\varphi'$ begins:
\be  V(0,\chi'|_{in}) =  \Lambda +
\frac{{(mm')}^2}{4|\alpha_\chi|},
          \lb{24y} \ee
and according to Eq.~(\ref{74z}), we have:
\be  V(0,\chi'|_{in}) =
         \Lambda + \frac{{81(vv')}^2}{256|\alpha_\chi|}. \lb{25y} \ee

\subsection{Inflation with the Higgs doublets of the $SM$ and $SM'$}

Our next step is an assumption that during the inflation the
scalar field $\chi$, which is the SM-triplet, decays into the two
Higgs doublets of the SM:
\be
      \chi \to \phi^{\dagger} + \phi.   \lb{32z} \ee
As a result, we have:
 \be \chi = a_d|\phi|^2, \lb{33z} \ee
where $\phi$ is the Higgs doublet field of the Standard Model. The
Higgs field $\phi$ also interacts directly with the mirror Higgs
field $\phi'$, according to the assumption of Ref.~\ct{12mw}:
\be V_{int} = \alpha_\phi
(\phi^{\dagger}\phi)({\phi'}^{\dagger}\phi'). \lb{30y} \ee
The Higgs field $\phi'$ has a time of the evolution and modifies
the shape of the barrier, so that at some value $\phi'_E$, it can
roll down the field $\phi$. This possibility, which we consider in
this paper, is given by the so-called Hybrid inflation scenario by
A.~Linde \ct{Lin}. Here we assume that the field $\phi$ begins the
inflation at the value $\phi|_{in}\simeq H_0$, where $H_0$ is the
Hubble rate (this is a result of Refs.~\ct{1S,2S}).

Using the action (\ref{15y}), given by GWU, we obtain near the
local ``false vacuum'' the following gravitational potential in
units $M_{Pl}^{red}=1$:
\be V(\phi,\,\,{\phi'}) \simeq \Lambda + \frac{\lambda}4 |\phi|^4
+ \frac{\lambda'}4|{\phi'}|^4 + \frac{a_\phi}4|\phi|^2|{\phi'}|^2,
\lb{34z} \ee
where $\lambda=12a_d^2m^2$  and ${\lambda}' = 12{a'_d}^2{m'}^2$
are self-couplings of the Higgs doublet fields $\phi$ and
${\phi}'$, respectively.

In the present investigation we considered only the results of
such an inflation, which corresponds to the assumption of the MPP,
that cosmological constant is zero (or almost zero) at both vacua:
at the "first vacuum" with the VEV $v_1 = 246$ GeV and at the
"second vacuum" with the VEV $v=v_2\sim 10^{18}$ GeV. If so, we
have the following conditions of the MPP (see section 4):
\be V_{eff} \left(\phi_{min 1}\right) = V_{eff} \left(\phi_{min
2}\right) = 0, \lb{20y} \ee
\be  \frac{\partial V_{eff}}{\partial |\phi|^2} \big|_ {\phi =
\phi_{min 1}} = \frac{\partial V_{eff}}{\partial |\phi|^2} \big|_
{\phi = \phi_{min 2}} = 0. \lb{36z} \ee
Being far from the Planck scale, we can present the following
expression for the low energy total effective Higgs potential:
 \be V_{eff} = - \frac{\mu^2}2 |\phi|^2 + \frac 14\lambda(\phi)|\phi|^4
 - \frac{{\mu'}^2}2 |{\phi'}|^2 + \frac 14\lambda'(\phi')|\phi'|^4
 + \frac 14\alpha_\phi(\phi, \phi') |\phi|^2 |{\phi'}|^2,
\lb{37z} \ee
where $\alpha_\phi (\phi, \phi')$ is a coupling for the
interaction of the ordinary Higgs field $\phi$ with the mirror
Higgs field $\phi'$. In Eq.~(\ref{37z}) the Higgs fields have
tachyonic masses $\mu,\mu'$, as usual.

According to the MPP, at the critical point of the phase diagram
of our theory, corresponding to the "second vacuum", we have:
\be \mu = \mu' \simeq 0, \quad \lambda (\phi_0) \simeq 0, \quad
{\lambda'}({\phi'}_0) \simeq 0, \lb{38z} \ee
and then
\be \alpha_\phi (\phi_0, {\phi'}_0) \simeq 0,\quad {\rm{if}}\quad
V_{eff}^{crit} (v_2) \simeq 0.  \lb{39z} \ee
At the critical point, corresponding to the first EW vacuum with
the VEV $v_1=246$ GeV, we also have $V_{eff}^{crit} (v_1) \simeq
0$, according to the MPP prediction of the existence of almost
degenerate vacua in the Universe. Then the tree-level Higgs
potential near vacua is
 \be V(\phi, \phi') = - \frac{\mu^2}2 |\phi|^2 + \frac 14\lambda |\phi|^4
 - \frac{{\mu'}^2}2 |{\phi'}|^2 + \frac 14\lambda' |\phi'|^4
 + \frac 14\alpha_\phi |\phi|^2 |{\phi'}|^2 + CC,
\lb{37'z} \ee
where CC is a constant, depending on the vacuum values of the
potential.

If $V(v_1,v'_1) = 0$, then CC (providing zero cosmological
constant) is:
\be CC = \frac{1}{4}\alpha_\phi(v_1v_1')^2,   \lb{40z} \ee
and we can present the tree-level Higgs potential by the following
expression:
\be V(\phi, \phi') = \frac 14\left(\lambda {(|\phi|^2 - v^2_1)}^2
+ \lambda' {(|\phi'|^2 - {v'}_1^2)}^2 + \alpha_\phi(|\phi'|^2 -
{v'}_1^2)(|\phi|^2 - v_1^2)\right),
                               \lb{41z} \ee
where in Eq.~(\ref{37'z}) we have:
\be \mu^2 = \lambda v_1^2 + \frac 12\alpha_\phi {v_1'}^2 \quad
{\rm {and}}\quad {\mu'}^2 = \lambda' {v'_1}^2  + \frac
12\alpha_\phi {v_1}^2.     \lb{41a} \ee
If  $\alpha_\phi = 0$, then we see from Eqs.~(\ref{40z}) that:
\be CC = 0,   \lb{41b} \ee
and the MPP-conditions (\ref{38z}) take place.

Eqs.~(\ref{41z}) determines VEVs:
\be v_1^2 = 2\frac{\alpha_\phi {\mu'}^2 - 2
\mu^2\lambda'}{\alpha_\phi^2 - 4 \lambda\lambda'} \quad {\rm{and}}
\quad {v'_1}^2 = 2\frac{\alpha_\phi {\mu}^2 - 2
{\mu'}^2\lambda}{\alpha_\phi^2 - 4 \lambda\lambda'}.
                                \lb{41c} \ee
If $\alpha_\phi = 0$, then:
\be v_1^2 = \frac{\mu^2}{\lambda} \quad {\rm{and}} \quad {v'_1}^2
= \frac{{\mu'}^2}{\lambda'},
                \lb{41d} \ee
and we have well-known relations for the usual Higgs model.

The potential (\ref{41z}) vanishes, when $\phi' = {\phi'}_0 =
v'_1$ and $\phi = {\phi}_0 = v_1$, recovering the ordinary or
mirror Standard Model, respectively.

At the end of inflation we have: $\phi' = {\phi'}_E$, and the
first vacuum value of the potential $V$ is:
\be V(v_1, {\phi'}_E) = \frac 14{\lambda'({\phi'}_E)}({\phi'}_E|^2
- {v'}_1^2)^2,
                                  \lb{42z} \ee
and
\be \frac{\partial V}{\partial |\phi|^2} \left|\begin{array}{l}
\phi = v_1\\
{\phi'} = {\phi'}_E\end{array}\right| = 0.
                       \lb{43z} \ee
This means that the end of inflation is given by the value:
\be \phi'_{end} = {\phi'}_E = v'_1 = \zeta v_1, \lb{44z} \ee
which coincides with the VEV \,$<{\phi'}>=v'_1$\,  of the field
${\phi'}$ at the first vacuum in the mirror world MW. Thus:
\be V(\phi,{\phi'}_E) = \frac14 \lambda {(|\phi|^2 - v^2_1)}^2,
\lb{45z} \ee
recovering the Standard Model with the first vacuum VEV
$v_1\approx 246 $ GeV.

\subsection{The calculation of the number of e-folds}

In this section we try to estimate the number of e-folds in our
model with two scalar fields, $\varphi$ and $\varphi'$. Here we
follow the calculations given by Ref.~\ct{FGH}.

Near the Planck scale the tree-level Higgs potential is given by
\be V = \frac{\lambda'}{4}(s^2 - v'^2)^2 + \frac{\lambda}{4}(h^2 -
v^2)^2 + \frac{\alpha_h}{4}(s^2 - v'^2)(h^2 - v^2),  \lb{46z} \ee
where $s=|\varphi'|$ and $h=|\varphi|$. In Eq.~(\ref{46z})
$s_0=v'$ and $h_0=v$ are VEVs of the Planck scale vacua ("false
vacua").

During inflation the mirror field $s$ rolls towards its minimum
$<s>$ and the mixing term between $h$ and $s$ will grow and lift
the false vacuum. The end of inflation is given by the point, at
which the false minimum disappears. Here we have a situation
similar to the hybrid potential by A.~Linde \ct{Lin}, in which the
rolling of $s$ triggers the waterfall field $h$. Then:
\be \frac{\partial V}{\partial s^2}|_{s=<s>,h=<h>} = 0, \lb{47z}
\ee
what gives:
\be  <s>^2 - v'^2 + \frac{\alpha_h}{2\lambda'}(<h>^2 - v^2) = 0,
                                           \lb{48z} \ee
and we have the following minimum of the tachyonic field $s$:
\be  <s>^2 = v'^2 - \frac{\alpha_h}{2\lambda'}(<h>^2 - v^2).
                               \lb{49z} \ee
According to (\ref{23y}), using (\ref{1z}), (\ref{71z}),
(\ref{72z}) and (\ref{74z}), we obtain:
\be <h>^2 =|v - \sigma_{in}|^2 = |v - g_W\chi_{in}|^2 = |v - \frac
38\frac {vv'}{\sqrt{2\alpha}}|^2,  \lb{50z} \ee
where $\alpha\equiv |\alpha_h|$ (Eqs.~(\ref{23y})-(\ref{25y}) tell
us that $\alpha_h < 0$). Then the renormalized potential can be
written as a function of $s$ (compare with Ref.~\ct{FGH}):
\be V_s = \frac{\lambda'}{4}(s^2 - <s>^2)^2 + \frac
14(\lambda_{eff} - \frac{{\alpha_h}^2}{2\lambda'})(<h>^2 - v^2)^2.
        \lb{51z} \ee
Here we neglected the $h^2$ term. In Eq.~(\ref{51z}) the value
$\lambda_{eff}\simeq 0.129$ is inferred from the Higgs mass
measurements (see Ref.~\ct{FGH}), and
\be  <s>^2 = \frac 1{2\lambda'}\left( {M_h'}^2 - \alpha_h(<h>^2 -
v^2)\right) = {v'}^2  + \frac{\alpha}{2\lambda'}(<h>^2 - v^2),
\lb{52z} \ee
where the mirror Higgs boson mass is given by the relation
${M_h'}^2 = 2\lambda'{v'}^2$.

Using Eq.~(\ref{50z}), we have:
\be <h>^2 - v^2 = \frac{\zeta v^3}{\alpha}\left(\frac{9}{128}\zeta
v - \frac 34 \sqrt{\frac{\alpha}{2}}\right).
                           \lb{53z} \ee
When the inflation begins, the "false vacua" disappear:
\be   <h>^2 = 0  \quad {\rm{and}} \quad <s>^2 = 0. \lb{55z} \ee
>From Eq.~(\ref{50z}) the condition $<h>^2 = 0$ (see
Eq.~(\ref{55z})) determines the value $\alpha$ (equal to
$\alpha_0$) at the beginning of inflation:
\be       \alpha_0 \simeq \frac{9}{128}\zeta^2v^2.   \lb{56z} \ee
Eqs.~(\ref{52z}) and (\ref{55z}) give:
\be  <s>^2 = \zeta^2 v^2  - \frac{\alpha}{2\lambda'}v^2 = 0,
 \lb{57z} \ee
and we obtain the following result:
\be  \alpha_0 = 2\lambda'\zeta^2.
                        \lb{58z} \ee
Then, according to Eqs.~(\ref{56z}) and (\ref{58z}), we have:
\be  \lambda' = \frac{9}{256}v^2.
                        \lb{59z} \ee
Taking into account the relations (\ref{51z}), (\ref{52z})  and
(\ref{53z}), we obtain:
\be   V_s = As^4 + Bs^2 + C,   \lb{58y} \ee
where
$$  A = \frac 14{\lambda'},$$
$$
B = - \frac 12{\lambda'}<s>^2,$$
\be   C = \frac 14{\lambda'}<s>^4 + \frac 9{512\alpha}(0.129 -
\frac{\alpha^2}{2\lambda'})\zeta^2v^4. \lb{59y} \ee
In cosmology the total number of e-folds in units $M_{Pl}^{red}=1$
(where $M_{Pl}^{red}\simeq 2.43\cdot 10^{18}$ GeV) is given by:
\be    N^* = \int_{t_{in}}^{t_{end}}H_0 dt,   \lb{60y} \ee
what means (see Ref.~\ct{FGH}):
\be    N^* = \int_{s_{in}}^{s_{end}}\frac{V_s}{V'_s}ds, \lb{61y}
\ee
where $s_{in}$ and $s_{end}$ are initial and end values of $s$
during inflation. The calculations give:
\be s_{in} = |\varphi'_{in}| = |v' - \sigma'_{in}| = |v' -
g'_w\chi'_{in}| = |v' - \frac{mv'}{4\sqrt{\alpha}}| = \zeta v |1 -
\frac 38\frac{v}{\sqrt{2\alpha}}|,  \lb{62y} \ee
and, according to relations (\ref{1z}), (\ref{71z}), (\ref{72z}),
(\ref{74z}), (\ref{33z}) and (\ref{44z}):
$$ s_{end} = |\varphi'_{end}| = |v' - \sigma'_{end}| = |v' -
g'_w\chi'_{end}| = |v' - g'_Wa'_d|\phi'_{end}|^2| $$ \be = |v' -
\frac 16\sqrt{\lambda'/3}|\phi'_{end}|^2| = |v' - \frac 16
\sqrt{\lambda'/3}(\zeta v_1)^2|, \lb{64y} \ee
or
\be s_{end} \simeq v'= \zeta v. \lb{65y} \ee
In Eq.~(\ref{61y}) we have:
\be V'_s = \frac{\partial V_s}{\partial s} = 2s \frac{\partial
V_s}{\partial s^2}.   \lb{66y} \ee
The model developed in the present paper gives (see
Eq.~(\ref{58y})):
\be  N^* = \frac 14\int_{s^2_{in}}^{s^2_{end}}\frac{(As^4 + Bs^2+
C)}{s^2(2As^2 + B)}ds^2.   \lb{67y} \ee
Now we can obtain from Eqs.~(\ref{59y}) the following estimations
of coefficients A, B, C:
\be   A=\frac{\lambda'}{4}, \quad B = - \frac{\lambda'}{2}a^2,
\quad {\rm{and}}\quad  C = \frac{\lambda'}{4}a^4 +
\frac{9}{512\alpha}(0.129 - \frac{\alpha^2}{2\lambda'})\zeta^2
v^4, \lb{68y} \ee
where
\be  a^2 = <s>^2.
                               \lb{69y} \ee
Finally, we obtain the following result:
\be    N^* = \frac 18 \left(x_{end} - x_{in}\right) +
\frac{B}{16A}\ln\frac{(x_{in} + B/2A)}{(x_{end} + B/2A)} +
\frac{C}{4B}\ln\left(\frac{x_{in}}{x_{end}}\frac{(x_{end} +
B/2A)}{(x_{in} + B/2A)} \right), \lb{70y} \ee
where $x = s^2$ ($x_{in,end} = s^2_{in,end}$, respectively).

Using (\ref{68y}), we have:
\be    N^* = \frac 18 \left(x_{end} - x_{in}\right) -
\frac{a^2}{8}\ln\frac{(x_{end} - a^2)}{(x_{in} - a^2)} +
\frac{C}{4B}\ln\left(\frac{x_{end}}{x_{in}}\frac{(x_{in} - a^2
B/2A)}{(x_{end} - a^2)} \right). \lb{71y} \ee
According to (\ref{55z}), the beginning of inflation gives:
\be   a^2 = 0, \lb{72y} \ee
and according to (\ref{68y}), we have:
\be \frac{B}{2A} = 0. \lb{73y} \ee
Then the number of e-folds is given by the following expression:
\be    N^* = \frac 18 (s^2_{end} - s^2_{in}). \lb{74y} \ee
Using the results (\ref{56z})-(\ref{59z}), (\ref{62y}) and
(\ref{65y}), we obtain:
\be  N^* \simeq \frac 14 \zeta v^2.
          \lb{75y} \ee
According to the "second vacuum" position (\ref{32u}), the VEV $v$
in units $M_{Pl}^{red} = 1$ is:
\be v \simeq \frac{3.5\cdot 10^{18}}{2.43\cdot 10^{118}}\simeq
1.44.                                 \lb{76y} \ee
Then
\be  N^* \simeq 0.52 \zeta.
          \lb{77y} \ee
Cosmological measurements give: $N^*\simeq 50-60$. Then
Eq.~(\ref{77y}) predicts:
\be \zeta \simeq  96   \lb{78y} \ee
-- for $N^*\simeq 50$, and
\be \zeta \simeq 115 \lb{79y} \ee
-- for $N^*\simeq 60$.

The values (\ref{78y}) and (\ref{79y}) for the MP parameter
$\zeta$ is in agreement with estimations obtained in
Refs.~\ct{10mw,11mw}, which predicted $\zeta\sim 100$. This means
that cosmology is consistent with our inflationary model.

According to Eq.~(\ref{56z}), at the beginning of inflation
$\alpha = \alpha_0$ is large:
\be  \alpha_0\sim 10^3,   \lb{80y} \ee
but $\lambda'$ (see Eq.~(\ref{59z})) is still small:
\be  \lambda'\sim 0.07.   \lb{81y} \ee
The subsequent reasonable results follow from the consideration of
the renormalization group equations (RGEs).

\section{Renormalization group equations for the ordinary and
mirror Higgs couplings}

In the effective Higgs theory the Higgs quartic coupling will be
modified to that of the Standard Models (SM and SM') as a result
of the mixing term. At high energies we have the one-loop RGEs,
given by Ref.~ \ct{MS}, but now we also include the mirror
s-field:
\be (4\pi)^2\beta_{\lambda} = (4\pi)^2\beta_{\lambda}^{(SM)} +
\frac 12 \alpha_h^2, \lb{1y} \ee
\be (4\pi)^2\beta_{\lambda'} = (4\pi)^2\beta_{\lambda'}^{(SM')} +
\frac 12 \alpha_h^2,
                  \lb{2y} \ee
\be  (4\pi)^2\beta_{\alpha_h} = 2\alpha_h^2 + 6\alpha_h(\lambda +
\lambda') + \frac 14\alpha_h(12y_t^2 - \frac 95 g_1^2 + 9 g_2^2 +
12{y'_t}^2 - \frac 95 {g'_1}^2 + 9 {g'_2}^2), \lb{3y} \ee
where
\be (4\pi)^2\beta_{\lambda}^{(SM)} = 12\lambda^2 + \lambda
(12y_t^2 - 9g_2^2 - 3g_1^2) + \frac{3}{4}g_1^4 + \frac{3}{2}g_1^2
g_2^2 + \frac{9}{4}g_2^4 - 12y_t^4, \lb{4y} \ee
and $(4\pi)^2\beta_{\lambda'}^{(SM')}$ is given by Eq.~(\ref{4y})
with replacements: $\lambda \to \lambda', y_t \to y'_t, g_{1,2}
\to g'_{1,2}$. Eqs.~(\ref{3y}) and (\ref{4y}) contain the
top-quark Yukawa coupling $y_t$, the $U_Y(1)$ coupling constant
$g_1$ and the $SU(2)$ coupling $g_2$. In order for this mechanism
not to change essentially the results of Refs.~ \ct{7,7a}, both
$\lambda'$ and $\alpha_h$ will need to be very small so the new
RGEs contributions will be minor.

We saw that near the Planck scale, parameters $\alpha_h$,
$\lambda$ and $\lambda'$ are very small, therefore the new RGE's
influence is not essential for the 2-, or 3-loop results (see
Refs.~\ct{7,7a}).

\section{Summary and Conclusions}

\begin{enumerate}

\item Using the Plebanski's formulation of gravity, we constructed
the Graviweak Unification (GWU) model, which is invariant under
the $G_{(GWU)} = Spin(4,4)$-group, isomorphic to the
$SO(4,4)$-group. Graviweak Unification is a model unifying gravity
with the weak $SU(2)$ gauge and Higgs fields.

\item Considering the Graviweak symmetry breaking, we have
obtained the following sub-algebras: $\mathfrak{\widetilde{g_1}}
=\mathfrak {sl}(2,C)^{(grav)}_L \oplus \mathfrak{su}(2)_L$ -- in
the ordinary world, and $\mathfrak{\widetilde{g'_1}} =
\mathfrak{{sl}}'(2,C)_R^{(grav)} \oplus \mathfrak {{su}'}(2)_R$ --
in the hidden world. These sub-algebras contain the self-dual
left-handed gravity in the OW, and the anti-self-dual right-handed
gravity in the MW. We showed, that finally at low energies we have
the Standard Model and the Einstein-Hilbert's gravity.

\item We discussed the existence of de Sitter solutions at the
early time of acceleration era of the Universe. It was shown that
in the ordinary world the VEV $v\sim 10^{18}$ GeV of "the false
vacuum" is given by the relation $v=R_0/3$. Here
$R_0={12}/{r_{dS}^2}$, where $r_{dS}$ is the radius of the
constant curvature of the de Sitter background space.

\item  We considered the Multiple Point Principle (MPP), which
postulates that the Nature has the Multiple Critical Point (MCP).
The MPP-model predicts the existence of several degenerate vacua
in the Universe, all having zero, or almost zero cosmological
constants.

\item We reviewed the Multiple Point Model (MPM) by D.L. Bennett
and H.B.Nielsen. We showed that the existence of two vacua into
the SM: the first one -- at the Electroweak scale ($v_1\simeq 246$
GeV), and the second one -- at the Planck scale ($v_2\sim 10^{18}$
GeV), was confirmed by calculations of the Higgs effective
potential in the 2-loop and 3-loop approximations. The
Froggatt-Nielsen's prediction of the top-quark and Higgs masses
was obtained in the assumption that there exist two degenerate
vacua into the SM.

\item In contrast to other theories of unification, we accepted an
assumption of the existence of visible and invisible (hidden)
sectors of the Universe. We gave arguments that modern
astrophysical and cosmological measurements lead to a model of the
Mirror World with a broken Mirror Parity (MP), in which the Higgs
VEVs of the visible and invisible worlds are not equal:
$\langle\phi\rangle=v, \quad \langle{\phi'}\rangle= v' \quad
{\rm{and}}\quad v\neq v'$. We considered a parameter
characterizing the violation of the MP: $\zeta = v'/v \gg 1$,
using the result: $\zeta \sim 100$ obtained by Z.~Berezhiani and
his collaborators.

\item In our model we showed that the action for gravitational and
$SU(2)$ Yang--Mills and Higgs fields, constructed in the ordinary
world (OW), has a modified duplication for the hidden (mirror)
world (MW) of the Universe.

\item We have developed a model of the Higgs inflation using the
GWU action, which contains a non-minimal coupling of the Higgs
field with gravity, suggested by F. Bezrukov and M. Shaposhnikov.
According to our model, a scalar field $\sigma$, being an
inflaton, starts trapped from the "false vacuum" of the Universe
at the value of the Higgs field's VEV $v =v_2 \sim 10^{18}$ GeV.
Considering the expansion of GWU Lagrangian in powers of small
values of $\sigma/v$, we get rid of the non-minimal coupling to
gravity by making the conformal transformation from the Jordan
frame to the Einstein frame.

\item We have used the Sidharth's prediction about the existence
of the discrete space-time at the Planck scale and his idea of
non-commutativity, which provides an almost zero cosmological
constant. This result was applied to our GWU model of the Higgs
inflation.

\item We assumed that during inflation inflaton $\sigma$ decays
into the two Higgs doublets of the SM: $\sigma\to \phi^\dagger
\phi$.

\item Taking into account the interaction between the initial
ordinary and mirror Higgs fields: $ \alpha_h( \phi^{\dagger}
\phi)( {\phi'} ^{\dagger}{\phi'}),$ we constructed a Hybrid model
of the Higgs inflation in the Universe. Such an interaction leads
to the emergence of the SM vacua at the EW scales: with the Higgs
boson VEVs $v_1\approx 246$ GeV -- in the OW, and $v'_1=\zeta v_1$
-- in the MW.

\item We have shown that our GWU model of the Higgs inflation is
in agreement with modern predictions of cosmology. We have
calculated the expression for a number of e-folds $N^*$ and have
obtained the following result for the MW parameter $\zeta$: $$
\zeta \simeq \frac{4N^*}{v^2}\simeq 100-115 \quad {\rm{for}} \quad
N^* \simeq 50-60,$$ in agreement with previous estimations
predicted $\zeta\sim 100$. This means that cosmology is consistent
with our GWU model of inflation.

\item We have calculated the  renormalization group equations
(RGEs) in the assumption that there exists the interaction between
the ordinary and mirror Higgs bosons. We discussed the possibility
of small values of parameters $\lambda'$ and $\alpha_h$ with aim
not to change essentially the results of the 2-, or 3-loop
calculations of the Higgs mass. We assumed that near the Planck
scale parameters $\alpha_h$, $\lambda$ and $\lambda'$ (according
to the MPP) are very small. Therefore, the influence of the
modified RGEs is not essential for the 2-, or 3-loop results of
Refs.~\ct{7,7a}.

\end{enumerate}

\section{Acknowledgments}

L.V.~Laperashvili greatly thanks
the B.M. Birla Science Centre (Hyderabad, India), and personally
Prof. B.G. Sidharth, for hospitality, collaboration and financial
support. C.R. Das acknowledges IOP for visiting scientist position.

\end{document}